\documentclass[
preprintnumbers,amsmath,amssymb,aps,prd]{revtex4}
\usepackage{amsthm,amscd,mathrsfs}
\usepackage{amsfonts,bm,latexsym}
\def\be{\begin{equation}}
\def\ee{\end{equation}}
\def\bea{\begin{eqnarray}}
\def\eea{\end{eqnarray}}
\def\ba{\begin{array}}
\def\ea{\end{array}}
\def\nn{\nonumber}
\def\p{\partial}
\def\sta{\sin\theta}
\def\cta{\cos\theta}
\def\sda{\sin^2\theta}
\def\cda{\cos^2\theta}

\begin{document}
\preprint{arXiv:0807.2114v3}

\title{Separability of the massive Dirac's equation in 5-dimensional Myers-Perry black hole geometry
and its relation to a rank-three Killing-Yano tensor}

\author{Shuang-Qing Wu \footnote{Electronic address: sqwu@phy.ccnu.edu.cn}}
\affiliation{College of Physical Science and Technology, Central China Normal University,
Wuhan, Hubei 430079, People's Republic of China}

\begin{abstract}
The Dirac equation for the electron around a five-dimensional rotating black hole with two different
angular momenta is separated into purely radial and purely angular equations. The general solution
is expressed as a superposition of solutions derived from these two decoupled ordinary differential
equations. By separating variables for the massive Klein-Gordon equation in the same spacetime
background, I derive a simple and elegant form for the St\"{a}ckel-Killing tensor, which can be
easily written as the square of a rank-three Killing-Yano tensor. I have also explicitly constructed
a symmetry operator that commutes with the scalar Laplacian by using the St\"{a}ckel-Killing tensor,
and the one with the Dirac operator by the Killing-Yano tensor admitted by the five-dimensional
Myers-Perry metric, respectively.
\end{abstract}

\pacs{04.50.Gh, 04.62.+v, 11.10.Kk, 03.65.Pm} 
\maketitle

\section{Introduction}

It is well known that the four-dimensional Kerr geometry \cite{RK} possesses a lot of miraculous
properties that not only can the geodesic Hamilton-Jacobi equation \cite{BC1} and the Klein-Gordon
scalar field equation \cite{BC1} be separated and decoupled into purely radial and purely angular
parts, but also the massless nonzero-spin field equations \cite{TME} as well as the equilibrium
equation for a stationary cosmic string \cite{FSZHCF}. These separability properties are shown to
be closely connected with the existence of an additional integral of motion associated with the
second order symmetric St\"{a}ckel-Killing tensor discovered in the Kerr metric by Carter \cite{BC1}.
The separation of the variables of Dirac's equation for massive fields in the Kerr geometry using
the Newman-Penrose formalism \cite{NP}, however, had only succeeded since Chandrasekhar's
remarkable work \cite{SC1}. Shortly after that, this result was extended by Page and other
people \cite{DKN} to the four-dimensional rotating charged Kerr-Newman black hole background.

As has been remarked by Chandrasekhar \cite{SC2}, the most striking feature of the Kerr metric
is the separability of all the standard wave equations in it. For some of these equations, their
separability has been understood as a consequence of the existence of certain tensor fields, which
have been found to be associated with a Killing spinor. Walker and Penrose \cite{WP} demonstrated
that the Carter's fourth constant can be constructed out of the Weyl spinor. Subsequently, the
separability of Dirac's equation has been explained by Carter and McLenaghan \cite{CM} in terms
of the existence of a Killing-Yano tensor, whose spinorial image is a two-index Killing spinor.
Physically, Killing-Yano tensors and operators constructed from them have been associated with
angular momentum. It has also been shown by a lot of people \cite{KS} that Killing-Yano tensors
and the Killing spinor play a crucial role in separation of variables for the Maxwell's equation
($s = 1$), Rarita-Schwinger's equation ($s = 3/2$), and the gravitational perturbation equation in
the Kerr geometry. The separation of various equations can be understood in terms of different
order differential operators that characterized the separation constants appeared in the separable
solutions. The differential operators characterizing separation constants \cite{HSF} are also
symmetry operators of the various field equations in question. The essential property that allows
the construction of such operators is the existence of a Killing-Yano tensor in the Kerr spacetime.
These results have been shown to hold for more general classes of type-D vacuum metrics; see Ref.
\cite{KMW} for a comprehensive review.

In recent years, higher-dimensional generalizations of the Kerr black hole and their properties have
attracted considerable attention \cite{HDBHs}, in particular, in the context of string theory, with
the discovery of the anti-de Sitter/conformal field theory (AdS/CFT) correspondence, and with the
advent of brane-world theories \cite{BW}, raising the possibility of direct observation of Hawking
radiation and of as probes of large spatial extra dimensions in future high energy colliders \cite{HEC}.
In the brane-world scenarios, our physical world is represented by a four-dimensional brane embedded
in the higher-dimensional bulk spacetime. Brane-world models of spacetimes with large extra dimensions
allow for the existence of higher-dimensional black holes whose geometry can be approximately described
by the classical solutions of vacuum Einstein equation, thus predicting the possibility of mini-black
hole production in a high energy factory. The metrics describing the isolated rotating black holes in
higher dimensions were first constructed by Myers and Perry \cite{MP} as the asymptotically flat
generalizations of the well-known four-dimensional Kerr vacuum solution. By introducing a nonzero
cosmological constant, Hawking, \textit{et al}. \cite{HHT} obtained the asymptotically nonflat
generalizations in five dimensions with two independent angular momenta and in higher dimensions
with just one nonzero angular momentum parameter. Further vacuum generalizations to all dimensions
have been made recently in \cite{GLPPCLP}. Quite recently, an exact charged generalization of the
Kerr-Newman solution in five dimensions was obtained in \cite{EMCS} within the framework of minimally
gauged supergravity theory. Other rotating charged black hole solutions in five-dimensional gauged
and ungauged supergravity were also obtained in \cite{CCLP,SUGBH,SEM,GodelBH}.

It is generally accepted that symmetries play a key role in the study of physical effects in the
gravitational fields of black holes. Initiated by the work of Frolov and his collaborators (see
\cite{VFR} for a review and references therein), recently there has been a resurgence of interest
\cite{FS,KF,KL,DKL,CGLP,VS,CLP,PD,KKAA,KKPV,FKK,HOY} in the study of ``hidden'' symmetry and
separation of variables properties of the Klein-Gordon scalar equation, the Hamilton-Jacobi
equation, and stationary strings \cite{Sstring} in higher dimensions \cite{CLHHOY}. Remarkably,
it was shown that the five-dimensional Myers-Perry \cite{MP} metric possesses a number of
miraculous properties similar to the Kerr metric. Namely, it allows the separation of variables
in the geodesic Hamilton-Jacobi equation and the separability of the massless Klein-Gordon scalar
field equation \cite{FS}. These properties are also intimately connected with the existence of
the second order St\"{a}ckel-Killing tensor \cite{FS} admitted by the five-dimensional Myers-Perry
black hole geometry. It was further demonstrated that this rank-two St\"{a}ckel-Killing tensor can
be constructed from its ``square root'', a rank-three Killing-Yano tensor \cite{KF}. Following the
procedure of Carter's construction \cite{BC2} in four dimensions, Frolov \textit{et al}. \cite{KF}
started from a potential 1-form to generate a rank-two conformal Killing-Yano tensor \cite{JL},
whose Hodge dual is just the expected Killing-Yano tensor. Subsequently, these results have
further been extended \cite{KF,KKPV,FKK,HOY} to general higher-dimensional rotating black
hole solutions with NUT charges \cite{GLPPCLP}.

However, less is known about the separability of Dirac's equation and other higher-spin fields
and its relation to the Killing-Yano tensor in higher dimensions \cite{MS,OY}. Therefore it is
important to investigate the separability of the fermion field equation and its relation to the
Killing-Yano tensor in higher-dimensional rotating black holes. In this paper, I will report
my past unpublished work (done in the October of 2004) on the separation of variables for a massive
Dirac equation in five-dimensional rotating Myers-Perry black holes with two unequal angular momenta
\cite{MP}. I will also present my recent construction of an explicit symmetry operator that commutes
with the standard Dirac operator, making use of the rank-three Killing-Yano tensor which can be
viewed as the square root of a rank-two symmetric St\"{a}ckel-Killing tensor. In addition, the
separated parts of a massive Klein-Gordon equation in the five-dimensional Myers-Perry background
are used to construct a simple and elegant expression of the St\"{a}ckel-Killing tensor. Note
that these symmetry operators are directly constructed from the separated solutions of the
Klein-Gordon equation and Dirac's equation in the background geometry considered here.

The outline of this paper goes as follows. In Sec. \ref{FFDE}, the action of the five-dimensional
gravity and fermions is given and the f\"{u}nfbein form of Dirac's equation is formulated. Using
Clifford algebra and the spinor representation of SO(4,1), I construct the spinor connection
1-form which is necessary for the fermion field equation in curved spacetime. Sec. \ref{DFEMY}
is devoted to dealing with the separation of variables of Dirac's equation in a five-dimensional
Myers-Perry black hole geometry. This section consists of three subsections. In Sec. \ref{FPMP},
a new form of the five-dimensional Myers-Perry metric is expressed in the Boyer-Lindquist coordinate
which admits an explicit construction of local orthonormal coframe 1-forms (pentad). A brief review
of the relevant symmetry properties of the Myers-Perry metric is also presented. In Sec. \ref{spcon},
the spinor connection is obtained by making use of the homomorphism between the SO(4,1) group and
its spinor representation which is derived from the Clifford algebra defined by the anticommutation
relations of the gamma matrices. In Sec. \ref{svDE}, the massive Dirac equation in five-dimensional
Myers-Perry black hole is separated into purely radial and purely angular equations. Sec. \ref{csosusy}
is also divided into three parts. In this section, the separated solutions of a massive Klein-Gordon
equation is used to construct a concise expression for the St\"{a}ckel-Killing tensor. From the
separated part of Dirac's equation, I also explicitly construct a first order symmetry operator that
commutes with the Dirac operator by using the rank-three Killing-Yano tensor. The last section \ref{CoRe}
is a brief summary of this paper and the related work under preparation. Possible applications of
this work to further research are given here. In Appendix A, the affine spin-connection 1-forms
are calculated by the first Cartan structure equation from the exterior differential of the pentad.
Appendix B displays the five-dimensional Myers-Perry metric in a manner similar to the Plebanski
solution \cite{PDle} in four dimensions.

\section{F\"{u}nfbein formalism of Dirac field equation
in 5-dimensional curved space} \label{FFDE}

It is well known that there exist two different but equivalent formalisms for the four-dimensional
gravity, namely, the orthonormal tetrad formalism \cite{CPCF} and the null-tetrad (Newman-Penrose)
formalism \cite{NP}. Dirac's equation in four dimensions was reformulated within the Newman-Penrose
formalism first by Chandrasekhar \cite{SC1} and then extended to the charged case by Page \cite{DKN}.
To my knowledge, a higher-dimensional generalization of the Newman-Penrose formalism was established
in \cite{HdNP,TypeD}, but no similar work was given for the Dirac equation, subject to the purpose
here. In absence of a similar Newman-Penrose formalism in five-dimensions, in this paper I will work
out the Dirac equation within the orthonormal pentad formalism \cite{d5KKD}. In a forthcoming paper
\cite{WuNP}, a seminull pentad formalism of the Dirac equation was constructed in the five-dimensional
relativity similar to the famous work of Chandrasekhar's \cite{SC1}. The Dirac equation has been shown
to be decoupled into purely radial and purely angular parts which agree with the results presented
here.

In curved background spacetime, the action of the five-dimension gravity and fermions is given by
\be
S =\int d^5x\sqrt{-g}\Big[ \frac{-\mathcal{R}}{16\pi}
+i\overline{\psi}\gamma^Ae_A^{~\mu}\big(\p_{\mu} +\Gamma_{\mu}\big)\psi
+i\mu_e\overline{\psi}\psi\Big] \, ,
\label{act}
\ee
where $\mathcal{R}$ is the five-dimensional curvature scalar of the metric $g_{\mu\nu}$, $\psi$ is a
four-component Dirac spinor, $\mu_e$ is the mass of the electron, $\Gamma_{\mu}$ is the spinor connection,
$e_A^{~\mu}$ is the f\"{u}nfbein (pentad), and $\gamma^A$'s are the five-dimensional gamma matrices.
My conventions are as follows: Latin letters $A, B$ denote local orthonormal (Lorentz) frame indices
$\{0, 1, 2, 3, 5\}$, while Greek letters $\mu, \nu$ run over five-dimensional spacetime coordinate
indices $\{t, r, \theta, \phi, \psi\}$. Units are used as $G = \hbar = c = 1$ throughout this paper.

The Dirac equation can be deduced from the action (\ref{act}) by variation with respect to the spinor
field as
\be
\big(\mathbb{H}_D +\mu_e\big)\Psi =
 \big[\gamma^Ae_A^{~\mu}(\p_{\mu} +\Gamma_{\mu}) +\mu_e\big]\Psi = 0 \, ,
\label{DE}
\ee
where the f\"{u}nfbein $e_A^{~\mu}$ and its inverse $e_{~\mu}^A$ are defined by the spacetime metric
$g_{\mu\nu} = \eta_{AB}e_{~\mu}^Ae_{~\nu}^B$ with $\eta_{AB} = diag (-1, 1, 1, 1, 1)$ being the flat
(Lorentz) metric tensor. For my purpose in this paper, I choose gamma matrices $\gamma^A$ obeying
the anticommutation relations (Clifford algebra)
\be
\big\{\gamma^A, \gamma^B\big\} \equiv \gamma^A\gamma^B +\gamma^B\gamma^A = 2\eta^{AB} \, ,
\label{Clifford}
\ee
and take an explicit representation of the Clifford algebra as follows:
\bea
&& \gamma^0 = i\Big(\ba{cc}
 0 & I \\
 I & 0 \ea \Big) \, , \qquad
\gamma^1 = i\Big(\ba{cc}
 0         & \sigma^3 \\
 -\sigma^3 & 0 \ea \Big) \, , \qquad
\gamma^2 = i\Big(\ba{cc}
 0         & \sigma^1 \\
 -\sigma^1 & 0 \ea \Big) \, , \qquad
\gamma^3 = i\Big(\ba{cc}
 0         & \sigma^2 \\
 -\sigma^2 & 0 \ea \Big) \, , \qquad
\gamma^5 = \Big(\ba{cc}
 I & 0 \\
 0 & -I \ea \Big) \, , \qquad
\label{GMr}
\eea
where $\sigma^i$'s are the Pauli matrices, and $I$ is a $2 \times 2$ identity matrix, respectively.

In order to derive the spinor connection 1-form $\Gamma = \Gamma_{\mu}dx^{\mu} \equiv \Gamma_Ae^A$,
I first compute the spin-connection 1-form $\omega_{AB} = \omega_{AB\mu}dx^{\mu} \equiv f_{ABC}e^C$
in the orthonormal frame, i.e., the 1-form (pentad) $e^A = e^A_{~\mu}dx^{\mu}$ satisfying the
torsion-free condition
\be
de^A +\omega^A_{~B}\wedge e^B = 0 \, , \qquad \omega_{AB} = \eta_{AC}\omega_{~B}^C = -\omega_{BA} \, .
\label{CFE}
\ee
To obtain the spinor connection 1-form $\Gamma$ from $\omega_{AB}$, I can make use of the homomorphism
between the SO(4,1) group and its spinor representation which is derived from the Clifford algebra
(\ref{Clifford}). The SO(4,1) Lie algebra is defined by the ten antisymmetric generators $\Sigma^{AB}
= [\gamma^A, \gamma^B]/(2i)$ which gives the spinor representation, and the spinor connection $\Gamma$
can be regarded as a SO(4,1) Lie-algebra-valued 1-form. Using the isomorphism between the SO(4,1) Lie
algebra and its spinor representation, i.e., $\Gamma_{\mu} = (i/4)\Sigma^{AB}\omega_{AB\mu} = (1/4)
\gamma^A\gamma^B\omega_{AB\mu}$, I can immediately construct the spinor connection 1-form
\be
\Gamma = \frac{1}{8}\big[\gamma^A, \gamma^B\big]\omega_{AB}
= \frac{1}{4}\gamma^A\gamma^B\omega_{AB} = \frac{1}{4}\gamma^A\gamma^Bf_{ABC}e^C \, .
\ee
Now in terms of the local differential operator $\p_A = e_A^{~\mu}\p_{\mu}$, the Dirac equation
(\ref{DE}) can be rewritten in the local Lorentz frame as
\be
\big[\gamma^A(\p_A +\Gamma_A) +\mu_e\big]\Psi = 0 \, ,
\label{DELF}
\ee
where $\Gamma_A = e_A^{~\mu}\Gamma_{\mu} = (1/4)\gamma^B\gamma^Cf_{BCA}$ is the component of the
spinor connection in the local Lorentz frame. Note that the five-dimensional Clifford algebra has
two different but reducible representations (they can differ by the multiplier of a $\gamma^5$ matrix).
It is usually assumed that fermion fields are in a reducible representation of the Clifford algebra.
In other words, one can work with the Dirac equation in a four-component spinor formalism like in
the four-dimensional case, and just needs to take the $\gamma^5$ matrix as the fifth basis vector
component.

\section{Dirac field equation in 5-dimensional
Myers-Perry black hole} \label{DFEMY}

In this section, I will present a new form for the five-dimensional Myers-Perry metric in the
Boyer-Lindquist coordinates. One major advantage of these coordinates is that it allows us to
construct a local orthonormal pentad with which the Dirac equation can be decoupled into purely
radial and purely angular parts.

\subsection{Metric of a 5-dimensional
Myers-Perry black hole} \label{FPMP}

The metric of a five-dimensional rotating black hole with two independent angular momenta was first
obtained by Myers and Perry \cite{MP} in 1986. The solution with a negative cosmological constant
was given by Hawking \textit{et al}. \cite{HHT} in 1999. The line element of the Myers-Perry metric
can be recast into an elegant form in the Boyer-Lindquist coordinates as
\bea
ds^2 &=& g_{\mu\nu}dx^{\mu}dx^{\nu} = \eta_{AB}e^A\otimes e^B \nn \\
 &=& -\frac{\Delta_r}{\Sigma}\big(dt -a\sda {}d\phi -b\cda {}d\psi\big)^2
 +\frac{\Sigma}{\Delta_r}dr^2 +\Sigma {}d\theta^2 \nn \\
 && +\frac{\sda\cda}{p^2\Sigma}\big[(b^2-a^2)dt +(r^2+a^2)a d\phi -(r^2+b^2)b d\psi\big]^2 \nn \\
 &&~ +\frac{1}{r^2p^2}\big[-abdt +(r^2+a^2)b\sda {}d\phi +(r^2+b^2)a\cda {}d\psi\big]^2 \, ,
\label{KMP}
\eea
where
\be
\Delta_r = (r^2+a^2)(r^2+b^2)/r^2 -2M \, , \qquad \Sigma = r^2 +p^2 \, , \qquad
 p = \sqrt{a^2\cda +b^2\sda} \, . \nn
\ee

The metric determinant for this spacetime is $\sqrt{-g} = r\Sigma\sta\cta$, and the
contra-invariant metric tensor can be read accordingly from
\bea
g^{\mu\nu}\p_{\mu}\p_{\nu} &=& \eta^{AB}\p_A\otimes\p_B
 = -\frac{(r^2+a^2)^2(r^2+b^2)^2}{r^4\Delta_r\Sigma}\Big(\p_t
 +\frac{a}{r^2+a^2}\p_{\phi} +\frac{b}{r^2+b^2}\p_{\psi}\Big)^2
 +\frac{\Delta_r}{\Sigma}\p_r^2 +\frac{1}{\Sigma}\p_{\theta}^2 \nn \\
 &&\qquad +\frac{\sda\cda}{p^2\Sigma}\Big[(a^2-b^2)\p_t +\frac{a}{\sda}\p_{\phi}
 -\frac{b}{\cda}\p_{\psi}\Big]^2 +\frac{1}{r^2p^2}\big(ab\p_t +b\p_{\phi}
 +a\p_{\psi}\big)^2 \, .
\eea

The Myers-Perry metric (\ref{KMP}) possesses three Killing vectors ($\p_t$, $\p_{\phi}$, and $\p_{\psi}$),
In addition, it also admits a rank-two symmetric St\"{a}ckel-Killing tensor \cite{FS}, which can be
written as the square of a rank-three Killing-Yano tensor \cite{KF}. The existence of such tensors
ensures the separation of variables in the geodesic Hamilton-Jacobi equation and the separability
of the massless Klein-Gordon scalar field equation \cite{FS}. In this paper, it will be shown that
the separability of Dirac's equation in this spacetime background is also closely associated with the
existence of the rank-three Killing-Yano tensor.

The spacetime metric (\ref{KMP}) is of Petrov type-D \cite{PJDS,TypeD}. It possesses a pair of real
principal null vectors $\{\mathbf{l}, \mathbf{n}\}$, a pair of complex principal null vectors
$\{\mathbf{m}, \bar{\mathbf{m}}\}$, and one real, spatial-like unit vector $\mathbf{k}$. Similar
to the four-dimensional Kerr black hole case, they can be constructed to be of Kinnersley-type
as follows:
\bea
\mathbf{l}^{\mu}\p_{\mu} &=& \frac{(r^2+a^2)(r^2+b^2)}{r^2\Delta_r}\Big(\p_t
 +\frac{a}{r^2+a^2}\p_{\phi} +\frac{b}{r^2+b^2}\p_{\psi}\Big) +\p_r \, , \nn \\
\mathbf{n}^{\mu}\p_{\mu} &=& \frac{(r^2+a^2)(r^2+b^2)}{2r^2\Sigma}\Big(\p_t
 +\frac{a}{r^2+a^2}\p_{\phi} +\frac{b}{r^2+b^2}\p_{\psi}\Big)
 -\frac{\Delta_r}{2\Sigma}\p_r \, , \nn \\
\mathbf{m}^{\mu}\p_{\mu} &=& \frac{1}{\sqrt{2}(r+ip)}\Big\{\p_{\theta} +i\frac{\sta\cta}{p}
 \Big[(a^2-b^2)\p_t +\frac{a}{\sda}\p_{\phi} -\frac{b}{\cda}\p_{\psi}\Big]\Big\} \, , \nn \\
\bar{\mathbf{m}}^{\mu}\p_{\mu} &=& \frac{1}{\sqrt{2}(r-ip)}\Big\{\p_{\theta} -i\frac{\sta\cta}{p}
 \Big[(a^2-b^2)\p_t +\frac{a}{\sda}\p_{\phi} -\frac{b}{\cda}\p_{\psi}\Big]\Big\} \, , \nn \\
\mathbf{k}^{\mu}\p_{\mu} &=& \frac{1}{rp}\big(ab\p_t +b\p_{\phi} +a\p_{\psi}\big) \, .
\eea
These vectors are geodesic and satisfy the following orthogonal relations
\be
\mathbf{l}^{\mu}\mathbf{n}_{\mu} = -1 \, ,  \qquad \mathbf{m}^{\mu}\bar{\mathbf{m}}_{\mu} = 1 \, ,
\qquad \mathbf{k}^{\mu}\mathbf{k}_{\mu} = 1 \, ,
\label{Ognr}
\ee
and all others are zero.

Here I briefly sketch the construction of a seminull pentad formalism in five dimensions, analogous
to the four-dimensional Newman-Penrose null-tetrad formalism. For the Myers-Perry black hole (\ref{KMP})
which has a topology of $S^3$ sphere, a most convenient seminull pentad should endow it with a pair of
real principal null vectors, a pair of complex principal null vectors, and a real unit vector, which
obey the above orthogonal relations (\ref{Ognr}). In terms of these vectors, the metric can be written
as
\be
ds^2 = -\mathbf{l}\otimes \mathbf{n} -\mathbf{n}\otimes \mathbf{l} +\mathbf{m}\otimes
 \bar{\mathbf{m}} +\bar{\mathbf{m}}\otimes \mathbf{m} +\mathbf{k}\otimes \mathbf{k} \, .
\ee
I shall refer to this seminull pentad formalism as the $2\bar{2}1$ formalism. In a forthcoming paper
\cite{WuNP}, the Dirac equation has been reformulated within this seminull pentad formalism and can
be decoupled into purely radial and purely angular parts in the five-dimensional Myers-Perry black
hole geometry.

On the other hand, for black ring solutions \cite{BRs} whose horizon topology is $S^2\times S^1$, the
most suitable seminull pentad formalism should possess a real, timelike unit vector $\mathbf{k}$ and
two pairs of complex principal null vectors $\{\mathbf{m}_1, \bar{\mathbf{m}}_1\}$ and $\{\mathbf{m}_2,
\bar{\mathbf{m}}_2\}$, satisfying the orthonormal relations: $\mathbf{k}^{\mu}\mathbf{k}_{\mu} =
\mathbf{m}_1^{\mu}\bar{\mathbf{m}}_{1\mu} = \mathbf{m}_2^{\mu}\bar{\mathbf{m}}_{2\mu} = 1$. Working
within such a $1\bar{2}\bar{2}$ formalism, the metric tensor can be written as $g_{\mu\nu} = -
\mathbf{k}_{\mu}\mathbf{k}_{\nu} +\mathbf{m}_{1\mu}\bar{\mathbf{m}}_{1\nu} +\bar{\mathbf{m}}_{1\mu}
\mathbf{m}_{1\nu} +\mathbf{m}_{2\mu}\bar{\mathbf{m}}_{2\nu} +\bar{\mathbf{m}}_{2\mu}
\mathbf{m}_{2\nu}$.

\subsection{Construction of covariant
spinor differential operator}
\label{spcon}

In the local Lorentz form of Dirac's equation, I need to find the local differential operator $\p_A
= e_A^{~\mu}\p_{\mu}$ and the spinor connection $\Gamma_A = e_A^{~\mu}\Gamma_{\mu}$ subject to the
Myers-Perry metric (\ref{KMP}). The orthonormal basis 1-vectors $\p_A $ dual to the pentad $e^{A}$
constructed in the Appendix Eq. (\ref{pentad}) are
\bea
&& \p_0 = \frac{(r^2+a^2)(r^2+b^2)}{r^2\sqrt{\Delta_r\Sigma}}\Big(\p_t
 +\frac{a}{r^2+a^2}\p_{\phi} +\frac{b}{r^2+b^2}\p_{\psi}\Big) \, , \nn \\
&& \p_1 = \sqrt{\frac{\Delta_r}{\Sigma}}\p_r \, , \qquad  \nn \\
&& \p_2 = \frac{1}{\sqrt{\Sigma}}\p_{\theta} \, , \nn \\
&& \p_3 = \frac{\sta\cta}{p\sqrt{\Sigma}}\Big[(a^2-b^2)\p_t
 +\frac{a}{\sda}\p_{\phi} -\frac{b}{\cda}\p_{\psi}\Big] \, , \nn \\
&& \p_5 = \frac{1}{rp}\big(ab\p_t +b\p_{\phi} +a\p_{\psi}\big) \, .
\eea

Taking use of the local Lorentz frame component $\Gamma_A$ and the gamma matrices with relation
$\gamma^5 = -i\gamma^0\gamma^1\gamma^2\gamma^3$, I get the composite expression
\bea
\gamma^A\Gamma_A &=& \frac{1}{4}\gamma^A\gamma^B\gamma^Cf_{BCA} \nn \\
&=& \gamma^1\sqrt{\frac{\Delta_r}{\Sigma}}\Big(\frac{\Delta_r^{\prime}}{4\Delta_r}
 +\frac{1}{2r} +\frac{r}{2\Sigma} \Big) ~+\gamma^2\frac{1}{\sqrt{\Sigma}}\Big[
 \frac{1}{2}\cot\theta -\frac{1}{2}\tan\theta -\frac{(a^2-b^2)\sta\cta}{2\Sigma}\Big] \nn \\
&& ~-\frac{(a^2-b^2)r\sta\cta}{2\Sigma^{3/2}p}\gamma^0\gamma^1\gamma^3
 ~-\frac{ab}{2r^2p}\gamma^0\gamma^1\gamma^5
 ~+\frac{p\sqrt{\Delta_r}}{2\Sigma^{3/2}}\gamma^0\gamma^2\gamma^3
 ~+\frac{ab}{2rp^2}\gamma^2\gamma^3\gamma^5 \nn \\ %
&=& \gamma^1\sqrt{\frac{\Delta_r}{\Sigma}}\Big(\frac{\Delta_r^{\prime}}{4\Delta_r}
 +\frac{1}{2r} +\frac{r -ip\gamma^5}{2\Sigma} \Big) ~+\gamma^2\frac{1}{\sqrt{\Sigma}}
 \Big[\frac{1}{2}\cot\theta -\frac{1}{2}\tan\theta \nn \\
&&~ -\frac{(a^2-b^2)\sta\cta}{2\Sigma p}\big(p +ir\gamma^5\big)\Big]
  ~+\frac{iab}{2r^2p^2}\gamma^0\gamma^1\big(r +ip\gamma^5\big) \, ,
\eea
where a prime denotes the partial differential with respect to the coordinates $r$ and $\theta$.

Combining this formula with the spinor differential operator
\bea
&& \gamma^A\p_A = \gamma^0\frac{(r^2+a^2)(r^2+b^2)}{r^2\sqrt{\Delta_r\Sigma}}\Big(\p_t
 +\frac{a}{r^2+a^2}\p_{\phi} +\frac{b}{r^2+b^2}\p_{\psi}\Big)
 ~+\gamma^1\sqrt{\frac{\Delta_r}{\Sigma}}\p_r ~+\gamma^2\frac{1}{\sqrt{\Sigma}}\p_{\theta} \nn \\
&&\qquad\quad  ~+\gamma^3\frac{\sta\cta}{p\sqrt{\Sigma}}\Big[(a^2-b^2)\p_t
 +\frac{a}{\sda}\p_{\phi} -\frac{b}{\cda}\p_{\psi}\Big]
 ~+\gamma^5\frac{1}{rp}\big(ab\p_t +b\p_{\phi} +a\p_{\psi}\big) \, ,
\eea
I find that the covariant Dirac differential operator in the local Lorentz frame is
\bea
\gamma^A(\p_A +\Gamma_A) &=& \gamma^0\frac{(r^2+a^2)(r^2+b^2)}{r^2\sqrt{\Delta_r\Sigma}}
 \Big(\p_t +\frac{a}{r^2+a^2}\p_{\phi} +\frac{b}{r^2+b^2}\p_{\psi}\Big) \nn \\
&& +\gamma^1\sqrt{\frac{\Delta_r}{\Sigma}}\Big(\p_r +\frac{\Delta_r^{\prime}}{4\Delta_r}
 +\frac{1}{2r} +\frac{r -ip\gamma^5}{2\Sigma}\Big) +\gamma^2\frac{1}{\sqrt{\Sigma}}
 \Big[\p_{\theta} +\frac{1}{2}\cot\theta -\frac{1}{2}\tan\theta \nn \\
&& -\frac{(a^2-b^2)\sta\cta}{2\Sigma p}\big(p +ir\gamma^5\big)\Big]
 +\gamma^3\frac{\sta\cta}{p\sqrt{\Sigma}}\Big[(a^2-b^2)\p_t
 +\frac{a}{\sda}\p_{\phi}  -\frac{b}{\cda}\p_{\psi}\Big] \nn \\
&& +\gamma^5\frac{1}{rp}\big(ab\p_t +b\p_{\phi} +a\p_{\psi}\big)
 +\frac{iab}{2r^2p^2}\gamma^0\gamma^1\big(r +ip\gamma^5\big) \, .
\eea

\subsection{Separation of variables
in Dirac equation} \label{svDE}

With the above preparation in hand, I am now ready to decouple the Dirac equation. Substituting the
above spinor differential operator into Eq. (\ref{DELF}), the Dirac equation in the five-dimensional
Myers-Perry metric reads
\bea
&& \Bigg\{ \gamma^0\frac{(r^2+a^2)(r^2+b^2)}{r^2\sqrt{\Delta_r\Sigma}}\Big(\p_t
 +\frac{a}{r^2+a^2}\p_{\phi} +\frac{b}{r^2+b^2}\p_{\psi}\Big)
 +\gamma^1\sqrt{\frac{\Delta_r}{\Sigma}}\Big(\p_r +\frac{\Delta_r^{\prime}}{4\Delta_r}
 +\frac{1}{2r} +\frac{r -ip \gamma^5}{2\Sigma}\Big) \nn \\
&&\qquad +\gamma^2\frac{1}{\sqrt{\Sigma}}\Big[\p_{\theta} +\frac{1}{2}\cot\theta
 -\frac{1}{2}\tan\theta -\frac{(a^2-b^2)\sta\cta}{2\Sigma p}
 i\gamma^5(r -ip\gamma^5)\Big] \nn \\
&&\qquad\quad +\gamma^3\frac{\sta\cta}{p\sqrt{\Sigma}}\Big[(a^2-b^2)\p_t
 +\frac{a}{\sda}\p_{\phi} -\frac{b}{\cda}\p_{\psi}\Big]
 +\gamma^5\frac{1}{rp}\big(ab\p_t +b\p_{\phi} +a\p_{\psi}\big) \nn \\
&&\qquad\qquad +\frac{iab}{2r^2p^2}\gamma^0\gamma^1\big(r +ip\gamma^5\big)
 +\mu_e \Bigg\}\Psi = 0 \, .
\eea
Multiplying $(r -ip\gamma^5)\sqrt{r +ip\gamma^5} = \sqrt{\Sigma(r -ip\gamma^5)}$ by the left
to the above equation, and after some lengthy algebra manipulations I finally obtain
\bea
&& \Bigg\{ \gamma^0\frac{(r^2+a^2)(r^2+b^2)}{r^2\sqrt{\Delta_r}}\Big(\p_t
 +\frac{a}{r^2+a^2}\p_{\phi} +\frac{b}{r^2+b^2}\p_{\psi}\Big)
 +\gamma^1\sqrt{\Delta_r}\Big(\p_r +\frac{\Delta_r^{\prime}}{4\Delta_r} +\frac{1}{2r}\Big) \nn \\
&&~ +\gamma^2\Big(\p_{\theta} +\frac{1}{2}\cot\theta -\frac{1}{2}\tan\theta \Big)
 +\gamma^3\frac{\sta\cta}{p}\Big[(a^2-b^2)\p_t
 +\frac{a}{\sda}\p_{\phi} -\frac{b}{\cda}\p_{\psi}\Big] \nn \\
&&~ +\Big(\frac{\gamma^5}{p} -\frac{i}{r}\Big)\big(ab\p_t +b\p_{\phi}
 +a\p_{\psi}\big) +\frac{iab}{2}\Big(\frac{1}{p^2} +\frac{1}{r^2}\Big)\gamma^0\gamma^1
 ~+\mu_e(r -ip\gamma^5) \Bigg\}\big(\sqrt{r +ip\gamma^5}\Psi\big) = 0 \, .
\label{presde}
\eea

At this stage, I assume that the spin-$1/2$ fermion fields are in a reducible representation of
the Clifford algebra, which can be taken as a four-component Dirac spinor. Applying the explicit
representation (\ref{GMr}) for the gamma matrices and adopting the following ansatz for the
separation of variables
\be
\sqrt{r +ip\gamma^5}\Psi = e^{i(m\phi+k\psi-\omega t)}\left(\ba{cl}
&\hspace*{-5pt} R_2(r)S_1(\theta) \\
&\hspace*{-5pt} R_1(r)S_2(\theta) \\
&\hspace*{-5pt} R_1(r)S_1(\theta) \\
&\hspace*{-5pt} R_2(r)S_2(\theta)
\ea\right) \, ,
\ee
I find that the Dirac equation in the five-dimensional Myers-Perry metric can be decoupled into
the purely radial parts
\bea
\Big[\sqrt{\Delta_r}D_r -i\frac{(r^2+a^2)(r^2+b^2)}{r^2\sqrt{\Delta_r}}\Big(\omega
  -\frac{ma}{r^2+a^2} -\frac{kb}{r^2+b^2}\Big)\Big]R_1
 = \Big[\lambda +i\mu_er -\frac{ab}{2r^2} -\frac{i}{r}\big(ab\omega
  -mb -ka\big)\Big]R_2 \, , \label{sdea} && \\
\Big[\sqrt{\Delta_r}D_r +i\frac{(r^2+a^2)(r^2+b^2)}{r^2\sqrt{\Delta_r}}\Big(\omega
  -\frac{ma}{r^2+a^2} -\frac{kb}{r^2+b^2}\Big)\Big]R_2
 = \Big[\lambda -i\mu_er -\frac{ab}{2r^2} +\frac{i}{r}\big(ab\omega
  -mb -ka\big)\Big]R_1 \, , \label{sdeb} &&
\eea
and the purely angular parts
\bea
&& \Big\{L_{\theta} +\frac{\sta\cta}{p}\Big[(a^2-b^2)\omega
  -\frac{ma}{\sda} +\frac{kb}{\cda}\Big]\Big\}S_1
 = \Big[\lambda +\mu_ep +\frac{ab}{2p^2} +\frac{1}{p}\big(ab\omega
  -mb -ka\big)\Big]S_2 \, , \label{sdec} \\
&& \Big\{L_{\theta} -\frac{\sta\cta}{p}\Big[(a^2-b^2)\omega
  -\frac{ma}{\sda} +\frac{kb}{\cda}\Big]\Big\}S_2
 = \Big[-\lambda +\mu_ep -\frac{ab}{2p^2} +\frac{1}{p}\big(ab\omega
   -mb -ka\big)\Big]S_1 \, , \label{sded}
\eea
in which I have introduced two operators
$$D_r = \p_r +\frac{\Delta_r^{\prime}}{4\Delta_r} +\frac{1}{2r} \, , \qquad
 L_{\theta} = \p_{\theta} +\frac{1}{2}\cot\theta -\frac{1}{2}\tan\theta \, . $$

Now the separated radial equation and the angular equation can be reduced into a master equation
containing only one component. For the radial part, I write them explicitly as
\bea
&& \frac{1}{r}\sqrt{\Delta_r}D_r\Big(r\sqrt{\Delta_r}D_rR_1\Big)
 +\Bigg\{\frac{(r^2+a^2)^2(r^2+b^2)^2}{r^4\Delta_r}\Big(\omega -\frac{ma}{r^2+a^2}
 -\frac{kb}{r^2+b^2}\Big)^2 \nn \\
&&\quad~ -\frac{(ab\omega -mb -ka)^2}{r^2} +2\mu_e(ab\omega -mb -ka)
 -\mu_e^2r^2 -\lambda^2 +\lambda\frac{ab}{r^2} \nn \\
&&\quad~  -\frac{a^2b^2}{4r^4} -\frac{\lambda +2i\mu_er
 +ab/(2r^2)}{\lambda r +i\mu_er^2 -ab/(2r) -i\big(ab\omega -mb
 -ka\big)}\Delta_rD_r +\Big[\frac{2i}{r} +i\frac{\Delta_r^{\prime}}{2\Delta_r} \nn \\
&&\qquad -\frac{\mu_er -iab/r^2 +\big(ab\omega -mb -ka\big)/r}{\lambda r
+i\mu_er^2 -ab/(2r) -i\big(ab\omega -mb -ka\big)}\Big]
 \frac{(r^2+a^2)(r^2+b^2)}{r^2}\Big(\omega -\frac{ma}{r^2+a^2} \nn \\
&&\qquad\quad -\frac{kb}{r^2+b^2}\Big) -\frac{2i}{r}\Big[(2r^2+a^2+b^2)\omega
 -ma -kb\Big] \Bigg\}R_1 = 0 \, ,
\eea
and
\bea
&& \frac{1}{r}\sqrt{\Delta_r}D_r\Big(r\sqrt{\Delta_r}D_rR_2\Big)
 +\Bigg\{\frac{(r^2+a^2)^2(r^2+b^2)^2}{r^4\Delta_r}\Big(\omega -\frac{ma}{r^2+a^2}
 -\frac{kb}{r^2+b^2}\Big)^2 \nn \\
&&\quad~ -\frac{(ab\omega -mb -ka)^2}{r^2} +2\mu_e(ab\omega -mb -ka)
 -\mu_e^2r^2 -\lambda^2 +\lambda\frac{ab}{r^2} \nn \\
&&\quad~ -\frac{a^2b^2}{4r^4} -\frac{\lambda -2i\mu_er
 +ab/(2r^2)}{\lambda r -i\mu_er^2 -ab/(2r) +i\big(ab\omega -mb
 -ka\big)}\Delta_rD_r -\Big[\frac{2i}{r} +i\frac{\Delta_r^{\prime}}{2\Delta_r} \nn \\
&&\qquad +\frac{\mu_er +iab/r^2 +\big(ab\omega -mb -ka\big)/r}{\lambda r
-i\mu_er^2 -ab/(2r) +i\big(ab\omega -mb -ka\big)}\Big]
 \frac{(r^2+a^2)(r^2+b^2)}{r^2}\Big(\omega -\frac{ma}{r^2+a^2} \nn \\
&&\qquad\quad -\frac{kb}{r^2+b^2}\Big) +\frac{2i}{r}\Big[(2r^2+a^2+b^2)\omega
 -ma -kb\Big] \Bigg\}R_2 = 0 \, .
\eea
From the above decoupled master equations, it is easy to see that they are more complicated than
the four-dimensional case derived by Chandrasekhar \cite{BC1}. As for the exact solution to these
equations, I expect they can be recast into the confluent form of Heun equation \cite{Heun}.

The case occurs similarly for the angular parts if I adopt $p = \sqrt{a^2\cda +b^2\sda}$ rather
than $\theta$ itself as a variable. Moreover, the angular part can be transformed into the radial
part if I make the replacement $p = ir$ in the case $M = 0$.

\section{Construction of symmetry operators in terms
of St\"{a}ckel-Killing and Killing-Yano tensors}
\label{csosusy}

In the last section, I have explicitly shown that Dirac's equation is separable in the five-dimensional
Myers-Perry black hole spacetime. In this section, I will demonstrate that this separability is intimately
related to the very existence of a rank-three Killing-Yano tensor admitted by the Myers-Perry metric.
Specifically speaking, I will construct a symmetry operator that commutes with the scalar Laplacian
by using the St\"{a}ckel-Killing tensor, and another one that commutes with the Dirac operator by the
Killing-Yano tensor. These symmetry operators are directly constructed from the separated solutions
of the Klein-Gordon equation and Dirac's equation.

\subsection{St\"{a}ckel-Killing tensor from the
separated solution of the Klein-Gordon equation}

In this subsection, I will present a simple and elegant form for the St\"{a}ckel-Killing tensor,
which can be easily written as the square of a rank-three Killing-Yano tensor. This symmetric
tensor is constructed from the separated solution of the Klein-Gordon scalar field equation
in the five-dimensional Myers-Perry metric.

To begin with, let us consider a massive Klein-Gordon scalar field equation
\be
\big(\Box -\mu_0^2\big)\Phi = \frac{1}{\sqrt{-g}}\p_{\mu}\big(\sqrt{-g}
 g^{\mu\nu}\p_{\nu}\Phi\big) -\mu_0^2\Phi = 0 \, ,
\ee
together with the ansatz $\Phi = R(r)S(\theta)e^{i(m\phi +k\psi -\omega t)}$. In the background
spacetime metric (\ref{KMP}), the massive scalar field equation reads
\bea
&&\Bigg\{ -\frac{(r^2+a^2)^2(r^2+b^2)^2}{r^4\Delta_r\Sigma}\Big(\p_t
 +\frac{a}{r^2+a^2}\p_{\phi} +\frac{b}{r^2+b^2}\p_{\psi}\Big)^2
 +\frac{1}{r\Sigma}\p_r\big(r\Delta_r\p_r\big)
 +\frac{1}{\Sigma\sta\cta}\p_{\theta}\big(\sta\cta\p_{\theta}\big) \nn \\
&&\qquad +\frac{\sda\cda}{p^2\Sigma}\Big[(a^2-b^2)\p_t +\frac{a}{\sda}\p_{\phi}
 -\frac{b}{\cda}\p_{\psi}\Big]^2 +\frac{1}{r^2p^2}\big(ab\p_t +b\p_{\phi}
 +a\p_{\psi}\big)^2 -\mu_0^2\Bigg \}\Phi = 0 \, .
\eea
Apparently, it can be separated into a radial part and an angular part,
\bea
&& \frac{1}{r}\p_r\big(r\Delta_r\p_rR\big)
 +\Big\{\frac{(r^2+a^2)^2(r^2+b^2)^2}{r^4\Delta_r}\Big(\omega
 -\frac{ma}{r^2+a^2} -\frac{kb}{r^2+b^2}\Big)^2 \nn \\
&&\qquad\qquad\qquad\quad -\frac{1}{r^2}\big(ab\omega -mb -ka\big)^2
 -\mu_0^2r^2 -\lambda^2 \Big\}R(r) = 0 \, , \label{srs} \\
&& \frac{1}{\sta\cta}\p_{\theta}\big(\sta\cta\p_{\theta}S\big)
 -\Big\{\frac{\sda\cda}{p^2}\Big[(a^2-b^2)\omega -\frac{ma}{\sda}
 +\frac{kb}{\cda}\Big]^2 \nn \\
&&\qquad\qquad\qquad\quad  +\frac{1}{p^2}\big(ab\omega -mb -ka\big)^2
 +\mu_0^2p^2 -\lambda^2\Big\}S(\theta) = 0 \, , \label{sra}
\eea
which can be transformed into the confluent form of Heun equation \cite{Heun,KL}.

Now from the separated Eqs. (\ref{srs}) and (\ref{sra}), I can construct a new dual field
equation as follows:
\bea
&&\Bigg\{ -p^2\frac{(r^2+a^2)^2(r^2+b^2)^2}{r^4\Delta_r\Sigma}\Big(\p_t
 +\frac{a}{r^2+a^2}\p_{\phi} +\frac{b}{r^2+b^2}\p_{\psi}\Big)^2
 +\frac{p^2}{r\Sigma}\p_r\big(r\Delta_r\p_r\big)
 -\frac{r^2}{\Sigma\sta\cta}\p_{\theta}\big(\sta\cta\p_{\theta}\big) \nn \\
&&\qquad -\frac{r^2\sda\cda}{p^2\Sigma}\Big[(a^2-b^2)\p_t +\frac{a}{\sda}\p_{\phi}
 -\frac{b}{\cda}\p_{\psi}\Big]^2 +\frac{p^2-r^2}{r^2p^2}\big(ab\p_t +b\p_{\phi}
 +a\p_{\psi}\big)^2 -\lambda^2\Bigg \}\Phi = 0 \, , \qquad
\eea
from which I can extract a second order symmetric tensor --- the so-called St\"{a}ckel-Killing
tensor
\bea
K^{\mu\nu}\p_{\mu}\p_{\nu} &=& -p^2\frac{(r^2+a^2)^2(r^2+b^2)^2}{r^4\Delta_r\Sigma}
 \Big(\p_t +\frac{a}{r^2+a^2}\p_{\phi} +\frac{b}{r^2+b^2}\p_{\psi}\Big)^2
 +p^2\frac{\Delta_r}{\Sigma}\p_r^2 -r^2\frac{1}{\Sigma}\p_{\theta}^2 \nn \\
&& -r^2\frac{\sda\cda}{p^2\Sigma}\Big[(a^2-b^2)\p_t
 +\frac{a}{\sda}\p_{\phi} -\frac{b}{\cda}\p_{\psi}\Big]^2
 +\frac{p^2-r^2}{r^2p^2}\big(ab\p_t +b\p_{\phi} +a\p_{\psi}\big)^2 \, .
\label{5dKT}
\eea
This symmetric tensor $K_{\mu\nu} = K_{\nu\mu}$ obeys the equation \cite{WP}
\be
K_{\mu\nu;\rho} +K_{\nu\rho;\mu} +K_{\rho\mu;\nu} = 0 \, ,
\label{Kte}
\ee
and is equivalent to the one found in \cite{FS,KF}, up to an ignorable constant.

In the local Lorentz coframe (\ref{pentad}), it has a simple, diagonal form $K_{AB} = diag
(-p^2, p^2, -r^2, -r^2, p^2-r^2)$. Using the St\"{a}ckel-Killing tensor, the above dual equation
can be written in a coordinate-independent form
\be
\big(\mathbb{K} -\lambda^2\big)\Phi = \frac{1}{\sqrt{-g}}\p_{\mu}\big(\sqrt{-g}
 K^{\mu\nu}\p_{\nu}\Phi\big) -\lambda^2\Phi = 0 \, .
\ee
It is obvious that the operator $\mathbb{K}$ commutes with the scalar Laplacian $\Box$. Working out the
commutator $[\mathbb{K}, \Box] = 0$ yields the Killing equation (\ref{Kte}) and the integrability condition
for the St\"{a}ckel-Killing tensor. These two operators have a classical analogue. In classical mechanics,
the scalar Laplacian $\Box$ corresponds to the Hamiltonian $g_{\mu\nu}\dot{x}^{\mu}\dot{x}^{\nu}$, while
the operator $\mathbb{K}$ to the Carter's constant $K_{\mu\nu}\dot{x}^{\mu}\dot{x}^{\nu}$. They are two
integrals of motion in addition to three constants from the Killing vector fields $\p_t$, $\p_{\phi}$,
and $\p_{\psi}$.

\subsection{Killing-Yano potential, (conformal)
Killing-Yano tensor, and St\"{a}ckel-Killing tensor}

Before constructing a first order symmetry operator that commutes with the Dirac operator, I first give
a brief review on the recent work \cite{KF,KKPV,FKK} about the construction of the St\"{a}ckel-Killing
tensor from the (conformal) Killing-Yano tensor.

Penrose and Floyd \cite{PF} discovered that the St\"{a}ckel-Killing tensor for the four-dimensional
Kerr metric can be written in the form $K_{\mu\nu} = f_{\mu\rho}f^{~\rho}_{\nu}$, where the skew-symmetric
tensor $f_{\mu\nu} = -f_{\nu\mu}$ is the Killing-Yano tensor \cite{KY,TK,DR} obeying the equation
$f_{\mu\nu;\rho} +f_{\mu\rho;\nu} = 0$. Using this object, Carter and McLenaghan \cite{CM} constructed
a first order symmetry operator that commutes with the massive Dirac operator. In the case of a
four-dimensional Kerr black hole ($D = 4$), the Killing-Yano tensor $f$ is of the rank two, its
Hodge dual $k = -{^*}f$ is a rank-two, antisymmetric, conformal Killing-Yano tensor \cite{JL}
obeying the equation
\be
k_{\alpha\beta;\gamma} +k_{\alpha\gamma;\beta} = \frac{1}{D-1}\big(g_{\alpha\beta}k^{\mu}_{~\gamma;\mu}
 +g_{\gamma\alpha}k^{\mu}_{~\beta;\mu} -2g_{\beta\gamma}k^{\mu}_{~\alpha;\mu}\big) \, .
\ee
This equation can be equivalently rewritten in the form proposed by Penrose \cite{PenE}
\be
\mathcal{P}_{\alpha\beta\gamma} = k_{\alpha\beta;\gamma}
 +\frac{1}{D-1}\big(g_{\beta\gamma}k^{\mu}_{~\alpha;\mu}
 -g_{\gamma\alpha}k^{\mu}_{~\beta;\mu}\big) = 0 \, .
\ee

A conformal Killing-Yano tensor $k$ is dual to the Killing-Yano tensor if and only if it is closed
$dk = 0$. This fact implies that there exists a potential 1-form $\hat{b}$ so that $k = d\hat{b}$.
Carter \cite{BC2} is the first one who found this potential to generate the Killing-Yano tensor for
the Kerr-Newman black hole.

Recently, these results have further been extended \cite{KF,KKPV,FKK,HOY} to general higher-dimensional
rotating black hole solutions. In the case of $D = 5$ dimensions, it was demonstrated \cite{KF} that
the rank-two St\"{a}ckel-Killing tensor can be constructed from its ``square root'', a rank-three, totally
antisymmetric Killing-Yano tensor. Following Carter's procedure \cite{BC2}, Frolov \textit{et al}.
\cite{KF} found a potential 1-form to generate a rank-two conformal Killing-Yano tensor \cite{JL},
whose Hodge dual $f = {^*}k$ is a rank-three Killing-Yano tensor.

Now restricting ourselves to the five-dimensional Myers-Perry metric, it is easy to check that the following
object constructed from the rank-three Killing-Yano tensor
\be
K_{\mu\nu} = -\frac{1}{2}f_{\mu\alpha\beta}f_{\nu}^{~\alpha\beta} \, ,
\ee
is just the rank-two, St\"{a}ckel-Killing tensor given in Eq. (\ref{5dKT}). The rank-three Killing-Yano
tensor $f$ obeying the equation
\be
f_{\alpha\beta\mu;\nu} +f_{\alpha\beta\nu;\mu} = 0 \, ,
\label{KYe}
\ee
can be taken as the Hodge dual $f = {^*}k$ of the 2-form $k = d\hat{b}$ via the following definition:
\be
f_{\alpha\beta\gamma} = ({^*}k)_{\alpha\beta\gamma} = \frac{1}{2}
 \sqrt{-g}\varepsilon_{\alpha\beta\gamma\mu\nu}k^{\mu\nu}\, .
\ee

The Killing-Yano potential found for the five-dimensional Myers-Perry metric is \cite{KF}
\be
2\hat{b} = \big(-r^2+a^2\cda +b^2\sda\big)dt +(r^2+a^2)a\sda {}d\phi +(r^2+b^2)b\cda {}d\psi \, ,
\ee
from which a conformal Killing-Yano tensor can be constructed
\be
k = d\hat{b} = r~e^0\wedge e^1 +p~e^2\wedge e^3 \, .
\ee
Adopting the convention $\varepsilon^{01235} = 1 = -\varepsilon_{01235}$ for the totally
antisymmetric tensor density $\varepsilon_{ABCDE}$, I find that the Killing-Yano tensor
is given by
\be
f = {^*}k = \big(-p~e^0\wedge e^1 +r~e^2\wedge e^3\big)\wedge e^5 \, .
\ee
In what follows, I shall show that this rank-three Killing-Yano tensor and its exterior
differential
\be
W = df = -4\frac{ab}{rp} ~e^0\wedge e^1\wedge e^2\wedge e^3 +4\frac{(a^2-b^2)
 \sta\cta}{p\sqrt{\Sigma}} ~e^0\wedge e^1\wedge e^2\wedge e^5
 +4\sqrt{\frac{\Delta_r}{\Sigma}} ~e^1\wedge e^2\wedge e^3\wedge e^5 \, ,
\ee
play a central role in constructing a first order symmetry operator that commutes with the
Dirac operator.

\subsection{Killing-Yano tensor from the
separated solution of the Dirac equation}

The last task is to construct a first order symmetry operator that commutes with the Dirac operator,
parallel to the work done by Carter and McLenaghan \cite{CM} in the case of a four-dimensional Kerr
black hole. I proceed to construct such an operator from the separated solutions (\ref{sdea}-\ref{sded})
of the Dirac equation. After some tedious algebra manipulations, I find that the following equation:
\bea
&& \Bigg\{ \gamma^0p\sqrt{\Delta_r}\Big(\p_r +\frac{\Delta_r^{\prime}}{4\Delta_r}
 +\frac{1}{2r}\Big) +\gamma^1p\frac{(r^2+a^2)(r^2+b^2)}{r^2\sqrt{\Delta_r}}\Big(\p_t
 +\frac{a}{r^2+a^2}\p_{\phi} +\frac{b}{r^2+b^2}\p_{\psi}\Big) \nn \\
&&\quad +\gamma^2(-r)\frac{\sta\cta}{p}\Big[(a^2-b^2)\p_t
 +\frac{a}{\sda}\p_{\phi} -\frac{b}{\cda}\p_{\psi}\Big]
 +\gamma^3r\Big(\p_{\theta} +\frac{1}{2}\cot\theta -\frac{1}{2}\tan\theta \Big) \nn \\
&&\quad~ -i\gamma^0\gamma^1\frac{\Sigma}{rp}\big(ab\p_t +b\p_{\phi}
 +a\p_{\psi}\big) +\frac{ab}{2}\Big(\frac{ip}{r^2} +\frac{\gamma^5r}{p^2}\Big)
  +\lambda\big(\gamma^5r -ip\big) \Bigg\}\big(\sqrt{r +ip\gamma^5}\Psi\big) = 0 \, ,
\eea
is a dual one to the Dirac equation (\ref{presde}). Expanding it, I get
\bea
&& \Bigg\{ \gamma^0p\sqrt{\frac{\Delta_r}{\Sigma}}\Big(\p_r
 +\frac{\Delta_r^{\prime}}{4\Delta_r} +\frac{1}{2r} +\frac{r -ip \gamma^5}{2\Sigma}\Big)
 +\gamma^1p\frac{(r^2+a^2)(r^2+b^2)}{r^2\sqrt{\Delta_r\Sigma}}\Big(\p_t
 +\frac{a}{r^2+a^2}\p_{\phi} \nn \\
&&\qquad +\frac{b}{r^2+b^2}\p_{\psi}\Big) +\gamma^2(-r)\frac{\sta\cta}{p\sqrt{\Sigma}}
 \Big[(a^2-b^2)\p_t +\frac{a}{\sda}\p_{\phi} -\frac{b}{\cda}\p_{\psi}\Big] \nn \\
&&\qquad~ +\gamma^3r\frac{1}{\sqrt{\Sigma}}\Big[\p_{\theta}
 +\frac{1}{2}\cot\theta -\frac{1}{2}\tan\theta
 -\frac{(a^2-b^2)\sta\cta}{2\Sigma p}i\gamma^5(r -ip\gamma^5)\Big] \nn \\
&&\qquad\quad -i\gamma^0\gamma^1(r+i\gamma^5p)\frac{1}{rp}
 \big(ab\p_t +b\p_{\phi} +a\p_{\psi}\big)
 +\frac{iab}{2rp} +\gamma^5\Big(\lambda -\frac{ab}{2r^2}
 +\frac{ab}{2p^2}\Big) \Bigg\}\Psi = 0 \, .
\eea
I do not hope $\gamma^5\lambda$ appears in the above equation, therefore I can multiply it
the $\gamma^5$ matrix by the left so as to rewrite it as
\bea
&& \Bigg\{ \gamma^5\gamma^0p\sqrt{\frac{\Delta_r}{\Sigma}}\Big(\p_r
 +\frac{\Delta_r^{\prime}}{4\Delta_r} +\frac{1}{2r} +\frac{r -ip \gamma^5}{2\Sigma}\Big)
 +\gamma^5\gamma^1p\frac{(r^2+a^2)(r^2+b^2)}{r^2\sqrt{\Delta_r\Sigma}}\Big(\p_t
 +\frac{a}{r^2+a^2}\p_{\phi} +\frac{b}{r^2+b^2}\p_{\psi}\Big) \nn \\
&&\qquad +\gamma^5\gamma^2(-r)\frac{\sta\cta}{p\sqrt{\Sigma}}
 \Big[(a^2-b^2)\p_t +\frac{a}{\sda}\p_{\phi} -\frac{b}{\cda}\p_{\psi}\Big] \nn \\
&&\qquad~ +\gamma^5\gamma^3r\frac{1}{\sqrt{\Sigma}}\Big[\p_{\theta}
 +\frac{1}{2}\cot\theta -\frac{1}{2}\tan\theta
 -\frac{(a^2-b^2)\sta\cta}{2\Sigma p}i\gamma^5(r -ip\gamma^5)\Big] \nn \\
&&\qquad\quad~ +\big(p\gamma^0\gamma^1 -r\gamma^2\gamma^3\big)\frac{1}{rp}
 \big(ab\p_t +b\p_{\phi} +a\p_{\psi}\big) +\frac{iab}{2rp}\gamma^5 +\lambda
 -\frac{ab}{2r^2} +\frac{ab}{2p^2} \Bigg\}\Psi = 0 \, ,
\eea
which can be put into an operator form
\be
\big(\mathbb{H}_f +\lambda\big)\Psi = 0 \, .
\ee

My final aim is to find the explicit expression for this symmetry operator $\mathbb{H}_f$. The
process to construct such an operator is more involved than the one to treat with the Dirac
operator $\mathbb{H}_D = \gamma^{\mu}\nabla_{\mu} = \gamma^{\mu}\big(\p_{\mu} +\Gamma_{\mu}\big)$.
However, the final result is extremely simple,
\be
\mathbb{H}_f = -\frac{1}{2}\gamma^{\mu}\gamma^{\nu}f^{\quad\rho}_{\mu\nu} \big(\p_{\rho}
 +\Gamma_{\rho}\big) -\frac{1}{64}\gamma^{\mu}\gamma^{\nu}\gamma^{\rho}\gamma^{\sigma}
 W_{\mu\nu\rho\sigma} \, .
\ee
Using the definition $W_{\mu\nu\rho\sigma} = -f_{\mu\nu\rho;\sigma} +f_{\nu\rho\sigma;\mu}
-f_{\rho\sigma\mu;\nu} +f_{\sigma\mu\nu;\rho}$ and the property of gamma matrices as well as
$f^{\rho}_{~\mu\nu;\rho} = 0$, I can also write the above operator in another form
\be
\mathbb{H}_f = -\frac{1}{2}\gamma^{\mu}\gamma^{\nu}f^{\quad\rho}_{\mu\nu}\nabla_{\rho}
 +\frac{1}{16}\gamma^{\mu}\gamma^{\nu}\gamma^{\rho}\gamma^{\sigma}f_{\mu\nu\rho;\sigma} \, .
\ee

The symmetry operator $\mathbb{H}_f$ constructed here has a lot of correspondences in different
contexts. It is the five-dimensional analogue to the nonstandard Dirac operator discovered by
Carter and McLenaghan \cite{CM} for the four-dimensional Kerr metric, which generates generalized
angular momentum quantum number. This operator corresponds to the nongeneric supersymmetric
generator in pseudoclassical mechanics \cite{Spcm}. Moreover, the 2-form field $L_{\mu\nu} =
f_{\mu\nu\rho} \dot{x}^{\rho}$ is parallel-propagated along the geodesic with a cotangent vector
$\dot{x}^{\mu}$, whose square is just the Carter's constant $-(1/2)L_{\mu\nu}L^{\mu\nu} =
K_{\mu\nu}\dot{x}^{\mu}\dot{x}^{\nu}$.

The existence of a rank-three Killing-Yano tensor is enough to explain the separability of the
Dirac equation in the five-dimensional Myers-Perry vacuum background. The operator $\mathbb{H}_f$
commutes with the standard Dirac operator $\mathbb{H}_D$. Expanding the commutation relation
$[\mathbb{H}_D, \mathbb{H}_f] = 0$ yields the Killing-Yano equation (\ref{KYe}) and the
integrability condition for the rank-three Killing-Yano tensor $f$.

\section{Concluding remarks}
\label{CoRe}

In this paper, I have investigated the separability of the Dirac equation in the five-dimensional
Myers-Perry metric and its relation to a rank-three Killing-Yano tensor. First, the field equation
for Dirac fermions in five-dimensional relativity is formulated in a f\"{u}nfbein formalism. The
spinor connection is constructed by the method of the Clifford algebra and its derived Lie algebra
SO(4,1). Second, an orthonormal pentad has been established for the Myers-Perry metric so that
one can easily deal with the Dirac equation in this background geometry. It is obviously shown that
Dirac's equation in the Myers-Perry metric can be separated into purely radial and purely angular
parts. Finally, from the separated solutions of the massive Klein-Gordon equation and Dirac's
equation, I have constructed two symmetry operators that commute with the scalar Laplacian and
the Dirac operator, respectively. A simple form for the St\"{a}ckel-Killing tensor was given so
that it can be easily understood as the square of a rank-three Killing-Yano tensor.

A comparison with the previously published work \cite{MS,OY} is made here. First, my work covers
the partial work done in \cite{MS} about the separation of variables of Dirac's equation in the
Myers-Perry black hole with two equal-magnitude angular momenta. In \cite{OY}, Dirac's equation
was separated in general higher-dimensional rotating Kerr-AdS-NUT black hole background. However,
because the role of angular momenta becomes less obvious, it seems difficult to directly apply
that work to study various properties of the Dirac field. What is more, symmetry operators that
(anti-)commute with the Dirac operator have not been found there. Although my work can serve as
a special case of that work and is announced later than it, the results presented in this paper can
be directly applied to study various aspects \cite{PDirac} of fermion fields, for example, Hawking
radiation \cite{SWH}, emission rates \cite{DNP}, quasinormal modes, instability \cite{InSta}, etc.
On the other hand, a nonstandard Dirac operator has been explicitly constructed here. In addition,
the representation of gamma matrices adopted in this paper is different from that used in \cite{OY}.

In a subsequent work \cite{WuNP}, I have constructed a seminull pentad formalism of the Dirac
equation in the five-dimensional relativity similar to the widely used null-tetrad formalism \cite{NP}.
The Dirac equation can be shown to be decoupled into purely radial and purely angular parts which
agree with the equations obtained here. The agreement assures that the Clifford-algebra formalism
is equivalent to my seminull pentad formalism. A paper based upon my previously unpublished notes
is being written.

Finally, the present work can be directly extended to the case of five-dimensional rotating black
holes with a nonzero cosmological constant \cite{HHT}. In another forthcoming paper, the present
work has been generalized to the charged case of five-dimensional rotating black holes in minimal
gauged and ungauged supergravity \cite{EMCS} with the inclusion of a Chern-Simons term. It is found
there that the usual Dirac equation can not be separated by variables. To ensure the separability
of fermion fields in this Einstein-Maxwell-Chern-Simons background geometry, one must include an
additional term in the action of spin-$1/2$ fields. A paper on this aspect is in preparation.

It is also an interesting question to investigate the separability of higher-spin field equations
(for example, Maxwell's equation and Rarita-Schwinger's equation) in the five-dimensional Myers-Perry
metric and its relation to a rank-three Killing-Yano tensor.

\bigskip
\textbf{Acknowledgments}:
The separation of variables for Dirac's equation in this paper is based upon a previously uncompleted
draft written at the end of 2004 when the work was supported by a start grant from Central China Normal
University. This work is currently supported by the Natural Science Foundation of China under Grant
No. 10675051.

\section*{Appendix A:~~ Pentad and connection 1-forms}
\def\theequation{A\arabic{equation}}
\setcounter{equation}{0}

The new form of the five-dimensional Myers-Perry metric (\ref{KMP}) admits the following local Lorentz
basis of 1-forms (pentad) defined as $e^A = e^A_{~\mu}dx^{\mu}$ orthonormal with respect to $\eta_{AB}$,
\bea
&& e^0 = \sqrt{\frac{\Delta_r}{\Sigma}}\big(dt -a\sda {}d\phi -b\cda {}d\psi\big) \, , \nn \\
&& e^1 = \sqrt{\frac{\Sigma}{\Delta_r}}dr \, , \nn \\
&& e^2 = \sqrt{\Sigma}d\theta \, , \nn \\
&& e^3 = \frac{\sta\cta}{p\sqrt{\Sigma}}\big[(b^2-a^2)dt
  +(r^2+a^2)a{} d\phi -(r^2+b^2)b{} d\psi\big] \, , \nn \\
&& e^5 = \frac{1}{rp}\big[-abdt +(r^2+a^2)b\sda {}d\phi +(r^2+b^2)a\cda {}d\psi\big] \, .
\label{pentad}
\eea
These coframe 1-forms are different from those used in \cite{AFA,SY}.

After some algebraic computations, I obtain the exterior differential of the coframe 1-forms as
\bea
&& de^0 = -\Big(\sqrt{\frac{\Delta_r}{\Sigma}}~\Big)_{,r} ~e^0\wedge e^1
 +\frac{(a^2-b^2)\sta\cta}{\Sigma^{3/2}} ~e^0\wedge e^2
 -\frac{2p\sqrt{\Delta_r}}{\Sigma^{3/2}} ~e^2\wedge e^3 \, , \nn \\
&& de^1 = \frac{(a^2-b^2)\sta\cta}{\Sigma^{3/2}} ~e^1\wedge e^2 \, , \nn \\
&& de^2 = \frac{r\sqrt{\Delta_r}}{\Sigma^{3/2}} ~e^1\wedge e^2 \, , \nn \\
&& de^3 = -\frac{2(a^2-b^2)r\sta\cta}{\Sigma^{3/2}p} ~e^0\wedge e^1
 +\frac{r\sqrt{\Delta_r}}{\Sigma^{3/2}} ~e^1\wedge e^3 +\frac{p}{\sta\cta}
 \Big(\frac{\sta\cta}{p\sqrt{\Sigma}}\Big)_{,\theta} ~e^2\wedge e^3 \, , \nn \\
&& de^5 = -\frac{2ab}{r^2p} ~e^0\wedge e^1 +\frac{1}{r}\sqrt{\frac{\Delta_r}{\Sigma}}
 ~e^1\wedge e^5 +\frac{2ab}{rp^2} ~e^2\wedge e^3
 -\frac{(a^2-b^2)\sta\cta}{p^2\sqrt{\Sigma}} ~e^2\wedge e^5 \, .
\eea

The spin-connection 1-form $\omega^A_{~B} = \omega^A_{~B\mu}dx^{\mu} = f^A_{~BC}e^C$ can be found
from the Cartan's first structure equation (\ref{CFE}) as follows:
\bea
&& \omega^0_{~1} = \Big(\sqrt{\frac{\Delta_r}{\Sigma}}~\Big)_{,r} ~e^0
 -\frac{(a^2-b^2)r\sta\cta}{\Sigma^{3/2}p} ~e^3 -\frac{ab}{r^2p} ~e^5 \, , \nn \\
&& \omega^0_{~2} = -\frac{(a^2-b^2)\sta\cta}{\Sigma^{3/2}} ~e^0
 -\frac{p\sqrt{\Delta_r}}{\Sigma^{3/2}} ~e^3 \, , \nn \\
&& \omega^0_{~3} = -\frac{(a^2-b^2)r\sta\cta}{\Sigma^{3/2}p} ~e^1
 +\frac{p\sqrt{\Delta_r}}{\Sigma^{3/2}} ~e^2 \, , \nn \\
&& \omega^0_{~5} =  -\frac{ab}{r^2p} ~e^1 \, , \nn \\
&& \omega^1_{~2} = -\frac{(a^2-b^2)\sta\cta}{\Sigma^{3/2}} ~e^1
-\frac{r\sqrt{\Delta_r}}{\Sigma^{3/2}} ~e^2 \, , \nn \\
&& \omega^1_{~3} = -\frac{(a^2-b^2)r\sta\cta}{\Sigma^{3/2}p} ~e^0
 -\frac{r\sqrt{\Delta_r}}{\Sigma^{3/2}} ~e^3 \, , \nn \\
&& \omega^1_{~5} =  -\frac{ab}{r^2p} ~e^0
 -\frac{1}{r}\sqrt{\frac{\Delta_r}{\Sigma}} ~e^5 \, , \nn \\
&& \omega^2_{~3} = -\frac{p\sqrt{\Delta_r}}{\Sigma^{3/2}} ~e^0 -\frac{p}{\sta\cta}
 \Big(\frac{\sta\cta}{p\sqrt{\Sigma}}\Big)_{,\theta}
 ~e^3 -\frac{ab}{rp^2} ~e^5 \, , \nn \\
&& \omega^2_{~5} = -\frac{ab}{rp^2} ~e^3
 +\frac{(a^2-b^2)\sta\cta}{p^2\sqrt{\Sigma}} ~e^5 \, , \nn \\
&& \omega^3_{~5} = \frac{ab}{rp^2} ~e^2 \, .
\eea
The local Lorentz frame component $\Gamma_A$ can be easily read from the spinor connection 1-form
$\Gamma \equiv \Gamma_Ae^A = (1/4)\gamma^A\gamma^B\omega_{AB}$ as
\bea
&& \Gamma_0 = \frac{1}{2}\Big[-\Big(\sqrt{\frac{\Delta_r}{\Sigma}}~\Big)_{,r}\gamma^0\gamma^1
 +\frac{(a^2-b^2)\sta\cta}{\Sigma^{3/2}}\gamma^0\gamma^2 \nn \\
&&\qquad~~ -\frac{(a^2-b^2)r\sta\cta}{\Sigma^{3/2}p}\gamma^1\gamma^3
 -\frac{ab}{r^2p}\gamma^1\gamma^5
 -\frac{p\sqrt{\Delta_r}}{\Sigma^{3/2}}\gamma^2\gamma^3\Big] \, , \nn \\
&& \Gamma_1 = \frac{1}{2}\Big[\frac{(a^2-b^2)r\sta\cta}{\Sigma^{3/2}p}\gamma^0\gamma^3
 +\frac{ab}{r^2p}\gamma^0\gamma^5 -\frac{(a^2-b^2)\sta\cta}{\Sigma^{3/2}}
 \gamma^1\gamma^2\Big] \, , \nn \\
&& \Gamma_2 = \frac{1}{2}\Big[-\frac{p\sqrt{\Delta_r}}{\Sigma^{3/2}}\gamma^0\gamma^3
 -\frac{r\sqrt{\Delta_r}}{\Sigma^{3/2}}\gamma^1\gamma^2
 +\frac{ab}{rp^2}\gamma^3\gamma^5\Big] \, , \nn \\
&& \Gamma_3 = \frac{1}{2}\Big[\frac{(a^2-b^2)r\sta\cta}{\Sigma^{3/2}p}\gamma^0\gamma^1
 +\frac{p\sqrt{\Delta_r}}{\Sigma^{3/2}}\gamma^0\gamma^2
 -\frac{r\sqrt{\Delta_r}}{\Sigma^{3/2}}\gamma^1\gamma^3 \nn \\
&&\qquad\quad -\frac{p}{\sta\cta}\Big(\frac{\sta\cta}{p\sqrt{\Sigma}}\Big)_{,\theta}
 \gamma^2\gamma^3 -\frac{ab}{rp^2}\gamma^2\gamma^5\Big] \, , \nn \\
&& \Gamma_5 = \frac{1}{2}\Big[\frac{ab}{r^2p}\gamma^0\gamma^1
 -\frac{1}{r}\sqrt{\frac{\Delta_r}{\Sigma}}\gamma^1\gamma^5 -\frac{ab}{rp^2}\gamma^2\gamma^3
 +\frac{(a^2-b^2)\sta\cta}{p^2\sqrt{\Sigma}}\gamma^2\gamma^5\Big] \, .
\label{Lsc}
\eea

\section*{Appendix B:~~ Plebanski-like form of the
$D = 5$ Myers-Perry metric}
\def\theequation{B\arabic{equation}}
\setcounter{equation}{0}

In some cases, it is more convenient to use $p = \sqrt{a^2\cda +b^2\sda}$ rather than $\theta$ as
the appropriate angle coordinate. In doing so, the five-dimensional Myers-Perry metric can be put
in a symmetric manner as follows:
\bea
ds^2 &=& -\frac{\Delta_r}{\Sigma}\Big[dt -\frac{(p^2-a^2)a}{b^2-a^2}d\phi
 -\frac{(p^2-b^2)b}{a^2-b^2}d\psi\Big]^2 +\frac{\Sigma}{\Delta_r}dr^2 \nn \\
 &&  +\frac{\Sigma}{\Delta_p}dp^2 +\frac{\Delta_p}{\Sigma}\Big[dt
 +\frac{(r^2+a^2)a}{b^2-a^2}d\phi +\frac{(r^2+b^2)b}{a^2-b^2}d\psi\Big]^2 \nn \\
 &&~ +\Big(\frac{ab}{rp}\Big)^2\Big[dt -\frac{(r^2+a^2)(p^2-a^2)}{(b^2-a^2)a}d\phi
 -\frac{(r^2+b^2)(p^2-b^2)}{(a^2-b^2)b}d\psi\Big]^2 \, ,
\eea
where
\be
\Delta_r = (r^2+a^2)(r^2+b^2)/r^2  -2M \, , \qquad
\Delta_p = -(p^2-a^2)(p^2-b^2)/p^2 \, , \qquad \Sigma = r^2 +p^2 \, . \nn
\ee

The following coordinate transformations:
\be
t = \tau +(a^2+b^2)u +a^2b^2v \, , \qquad \phi = a(u +b^2v) \, , \qquad
\psi = b(u +a^2v) \, ,
\ee
sends the metric to a Plebanski-like form \cite{PDle}
\be
ds^2 = -\frac{\Delta_r}{\Sigma}\big(d\tau +p^2du\big)^2 +\frac{\Sigma}{\Delta_r}dr^2
 +\frac{\Sigma}{\Delta_p}dp^2 +\frac{\Delta_p}{\Sigma}\big(d\tau -r^2du\big)^2
 +\Big(\frac{ab}{rp}\Big)^2\big[d\tau +(p^2-r^2)du -r^2p^2dv\Big]^2 \, ,
\ee
in which the role of angular momenta becomes less clear.



\begin{thebibliography}{99}

\bibitem{RK}
R.P. Kerr, Phys. Rev. Lett. \textbf{11}, 237 (1963).

\bibitem{BC1}
B. Carter, Phys. Rev. \textbf{174}, 1559 (1968).

\bibitem{TME}
S.A. Teukolsky, Phys. Rev. Lett. \textbf{29}, 1114 (1972);
Astrophys. J. \textbf{185}, 635 (1973);
W.G. Unruh, Phys. Rev. Lett. \textbf{31}, 1265 (1973).

\bibitem{FSZHCF}
V.P. Frolov, V.D. Skarzhinsky, A.I. Zel'nikov, and O. Heinrich, Phys. Lett. B \textbf{224}, 255 (1989);
B. Carter and V.P. Frolov, Class. Quantum Grav. \textbf{6}, 569 (1989).

\bibitem{NP}
E.T. Newman and R. Penrose, J. Math. Phys. \textbf{3}, 566 (1962); \textbf{4}, 998(E) (1963).

\bibitem{SC1}
S. Chandrasekhar, Proc. R. Soc. Lond. A \textbf{349}, 571 (1976); \textbf{350}, 565(E) (1976).

\bibitem{DKN}
D.N. Page, Phys. Rev. D \textbf{14}, 1509 (1976);
C.H. Lee, Phys. Lett. B \textbf{68}, 152 (1977);
I. Semiz, Phys. Rev. D \textbf{46}, 5414 (1992).

\bibitem{SC2}
S. Chandrasekhar, \textit{The Mathematical Theory of Black Holes} (Clarendon, Oxford, 1983).

\bibitem{WP}
M. Walker and R. Penrose, Commun. Math. Phys. \textbf{18}, 265 (1970);
L.P. Hughston, R. Penrose, P. Sommers, and M. Walker, Commun. Math. Phys. \textbf{27}, 303 (1972);
L.P. Hughston and P. Sommers, Commun. Math. Phys. \textbf{32}, 147 (1973);
N.M.J. Woodhouse, Commun. Math. Phys. \textbf{44}, 9 (1975).

\bibitem{CM}
B. Carter and R.G. McLenaghan, Phys. Rev. D \textbf{19}, 1093 (1979);
in \textit{Proceedings of the 2nd Marcel Grossmann Conference on General Relativity}, Trieste,
1979, edited by R. Ruffini (North-Holland, Amsterdam, 1982), pp. 575;
R.G. McLenaghan and Ph. Spindel, Phys. Rev. D \textbf{20}, 409 (1979);
N. Kamran and R.G. McLenaghan, Phys. Rev. D \textbf{30}, 357 (1984);
J. Math. Phys. \textbf{25}, 1019 (1984);
Lett. Math. Phys. \textbf{7}, 381 (1983);
E.G. Kalnins, W. Miller, and G.C. Williams, J. Math. Phys. \textbf{27}, 1893 (1986);
E.G. Kalnins and G.C. Williams, J. Math. Phys. \textbf{31}, 1739 (1990);
R.G. McLenaghan, S.N. Smith and D.M. Walker, Proc. R. Soc. Lond. A \textbf{456}, 2629 (2000).

\bibitem{KS}
E.G. Kalnins, W. Miller, and G.C. Williams, J. Math. Phys. \textbf{30}, 2360 (1989);
G.F. Torres del Castillo, J. Math. Phys. \textbf{29}, 971 (1988);
\textbf{30}, 2614 (1989);
\textbf{27}, 1583 (1986);
Proc. R. Soc. Lond. A \textbf{400}, 119 (1985);
G. Silva-Ortigoza, J. Math. Phys. \textbf{36}, 6929 (1995).

\bibitem{HSF} 
G.F. Torres del Castillo, J. Math. Phys. \textbf{29}, 2078 (1988);
E.G. Kalnins, W. Miller, and G.C. Williams, J. Math. Phys. \textbf{30}, 2925 (1989).

\bibitem{KMW}
E.G. Kalnins, W. Miller Jr, and G.C. Williams, Phil. Trans. R. Lond. A \textbf{340}, 337 (1992).

\bibitem{HDBHs}
G.T. Horowitz, gr-qc/0507080;
N.A. Obers, arXiv:0802.0519 [hep-th];
R. Emparan and H.S. Reall, arXiv:0801.3471 [hep-th];
B. Kleihaus, J. Kunz, and F. Navarro-Lerida, AIP Conf. Proc. \textbf{977}, 94 (2008).

\bibitem{BW}
N. Arkani-Hamed, S. Dimopoulos, and G. Dvali, Phys. Lett. B \textbf{429}, 263 (1998);
I. Antoniadis, N. Arkani-Hamed, S. Dimopoulos, and G. Dvali, Phys. Lett. B \textbf{436}, 257 (1998);
N. Kaloper, J. March-Russell, G.D. Starkman, and M. Trodden, Phys. Rev. Lett. \textbf{85}, 928 (2000);
L. Randall and R. Sundrum, Phys. Rev. Lett. \textbf{83}, 3370 (1999);
\textbf{83}, 4690 (1999).

\bibitem{HEC}
S. Dimopoulos and G. Landsberg, Phys. Rev. Lett. \textbf{87}, 161602 (2001);
S.B. Giddings and S. Thomas, Phys. Rev. D \textbf{65}, 056010 (2002).

\bibitem{MP}
R.C. Myers and M.J. Perry, Ann. Phys. (N.Y.) \textbf{172}, 304 (1986).

\bibitem{HHT}
S.W. Hawking, C.J. Hunter, and M.M. Taylor-Robinson, Phys. Rev. D \textbf{59}, 064005 (1999).

\bibitem{GLPPCLP}
G.W. Gibbons, H. L\"{u}, D.N. Page, and C.N. Pope, Phys. Rev. Lett. \textbf{93}, 171102 (2004);
J. Geom. Phys. \textbf{53}, 49 (2005);
W. Chen, H. L\"{u}, and C.N. Pope, Class. Quantum Grav. \textbf{23}, 5323 (2006);
Nucl. Phys. B \textbf{762}, 38 (2007);
T. Houri, T. Oota, and Y. Yasui, Phys. Lett. B \textbf{666}, 391 (2008).

\bibitem{EMCS}
Z.W. Chong, M. Cveti\v{c}, H. L\"{u}, and C.N. Pope, Phys. Rev. Lett. \textbf{95}, 161301 (2005).

\bibitem{CCLP}
Z.W. Chong, M. Cveti\v{c}, H. L\"{u}, and C.N. Pope, Phys. Rev. D \textbf{72}, 041901(R) (2005);
Phys. Lett. B \textbf{644}, 192 (2007);
J.W. Mei and C.N. Pope, Phys. Lett. B \textbf{658}, 64 (2007);
H. L\"{u}, J.W. Mei, and C.N. Pope, Nucl. Phys. B (in press), arXiv:0804.1152 [hep-th];
arXiv:0806.2204 [hep-th].

\bibitem{SUGBH}
J.C. Breckenridge, R.C. Myers, A.W. Peet, and C. Vafa, Phys. Lett. B \textbf{391}, 93 (1997);
D. Klemm and W.A. Sabra, Phys. Lett. B \textbf{503}, 147 (2001);
H.K. Kunduri, J. Lucietti, and H.S. Reall, J. High Energy Phys. 04 (2006) 036.

\bibitem{SEM}
M. Cveti\v{c}, H. L\"{u}, and C.N. Pope, Phys. Rev. D \textbf{70}, 081502(R) (2004);
Phys. Lett. B \textbf{598}, 273 (2004).

\bibitem{GodelBH}
E.G. Gimon and A. Hashimoto, Phys. Rev. Lett. \textbf{91}, 021601 (2003);
S.Q. Wu, Phys. Rev. Lett. \textbf{100}, 121301 (2008).

\bibitem{VFR}
V. Frolov, Nucl. Phys. B (Proc. Suppl.) \textbf{127}, 45 (2004);
Prog. Theor. Phys. Suppl. \textbf{172}, 210 (2008);
V.P. Frolov and D. Kubiznak, Class. Quantum Grav. \textbf{25}, 154005 (2008).

\bibitem{FS}
V.P. Frolov and D. Stojkovi\'{c}, Phys. Rev. D \textbf{67}, 084004 (2003);
\textbf{68}, 064011 (2003).

\bibitem{KF}
V.P. Frolov and D. Kubiz\v{n}\'{a}k, Phys. Rev. Lett. \textbf{98}, 011101 (2007);
D. Kubiz\v{n}\'{a}k and V.P. Frolov, Class. Quantum Grav. \textbf{24}, F1 (2007).

\bibitem{KL}
H.K. Kunduri and J. Lucietti, Phys. Rev. D \textbf{71}, 104021 (2005).

\bibitem{DKL}
P. Davis, H.K. Kunduri, and J. Lucietti, Phys. Lett. B \textbf{628}, 275 (2005);
H.K. Kunduri and J. Lucietti, Nucl. Phys. B \textbf{724}, 343 (2005).

\bibitem{CGLP}
Z.W. Chong, G.W. Gibbons, H. L\"{u}, and C.N. Pope, Phys. Lett. B \textbf{609}, 124 (2005).

\bibitem{VS}
M. Vasudevan and K.A. Stevens, Phys. Rev. D \textbf{72}, 124008 (2005);
M. Vasudevan, K.A. Stevens, and D.N. Page, Class. Quantum Grav. \textbf{22}, 339 (2005);
\textbf{22}, 1469 (2005).

\bibitem{CLP}
W. Chen, H. L\"{u}, and C.N. Pope, J. High Energy Phys. 04 (2006) 008;
P. Davis, Class. Quantum Grav. \textbf{23}, 3607 (2006).

\bibitem{PD}
P. Davis, Class. Quant. Grav. \textbf{23}, 6829 (2006).

\bibitem{KKAA}
P. Krtous, Phys. Rev. D \textbf{76}, 084035 (2007);
P. Connell, V.P. Frolov, and D. Kubiznak, Phys. Rev. D \textbf{78}, 024042 (2008).

\bibitem{KKPV}
D.N. Page, D. Kubiz\v{n}\'{a}k, M. Vasudevan, and P. Krtou\v{s}, Phys. Rev. Lett. \textbf{98}, 061102 (2007);
P. Krtou\v{s}, D. Kubiz\v{n}\'{a}k, D.N. Page, and V.P. Frolov, J. High Energy Phys. 02 (2007) 004;
P. Krtous, D. Kubiznak, D.N. Page and M. Vasudevan, Phys. Rev. D \textbf{76}, 084034 (2007);
D. Kubiznak and P. Krtous, Phys. Rev. D \textbf{76}, 084036 (2007).

\bibitem{FKK}
V.P. Frolov, P. Krtou\v{s}, and D. Kubiz\v{n}\'{a}k, J. High Energy Phys. 02 (2007) 005;
A. Sergyeyev and P. Krtous, Phys. Rev. D \textbf{77}, 044033 (2008);
P. Krtous, V.P. Frolov, and D. Kubiznak, Phys. Rev. D \textbf{78} (in press), arXiv:0804.4705 [hep-th].

\bibitem{HOY}
T. Houri, T. Oota, and Y. Yasui, Phys. Lett. B \textbf{656}, 214 (2007);
Y. Yasui, Int. J. Mod. Phys. \textbf{23}, 2169 (2008);
T. Houri, T. Oota, and Y. Yasui, J. Phys. A: Math. Theor. \textbf{41}, 025204 (2008);
arXiv:0805.3877 [hep-th].

\bibitem{Sstring}
V.P. Frolov and K.A. Stevens, Phys. Rev. D \textbf{70}, 044035 (2004);
D. Kubiznak and V.P. Frolov, J. High Energy Phys. 02 (2008) 007;
H. Ahmedov and A.N. Aliev, Phys. Rev. D \textbf{78} (in press), arXiv:0805.1594 [hep-th].

\bibitem{CLHHOY}
W. Chen and H. L\"{u}, Phys. Lett. B \textbf{658}, 158 (2008);
N. Hamamoto, T. Houri, T. Oota, and Y. Yasui, J. Phys. A: Math. Theor. \textbf{40}, F177 (2007).

\bibitem{BC2}
B. Carter, J. Math. Phys. \textbf{28}, 1535 (1987).

\bibitem{JL}
J. Jezierski, Class. Quantum Grav. \textbf{14}, 1679 (1997);
\textbf{19} (2002) 4405;
J. Jezierski and M. Lukasik, Class. Quantum Grav. \textbf{23}, 2895 (2006).

\bibitem{MS}
K. Murata and J. Soda, Class. Quantum Grav. \textbf{25}, 035006 (2008).

\bibitem{OY}
T. Oota and Y. Yasui, Phys. Lett. B \textbf{659}, 688 (2008).

\bibitem{PDle}
J.F. Plebanski, Ann. Phys. (N.Y.) \textbf{90}, 196 (1975);
J.F. Plebanski and M. Demianski, Ann. Phys. (N.Y.) \textbf{98}, 98 (1976).

\bibitem{CPCF}
J.M. Cohen and R.T. Powers, J. Math. Phys. \textbf{7}, 238 (1966);
Commun. Math. Phys. \textbf{86}, 69 (1982).

\bibitem{HdNP}
V. Pravda, A. Pravdova, A. Coley, and R. Milson, Class. Quantum Grav. \textbf{21}, 2873 (2004);
\textbf{24}, 1691(E) (2007);
M. Ortaggio, V. Pravda, and A. Pravdova, Class. Quantum Grav. \textbf{24}, 1657 (2007).

\bibitem{TypeD}
A. Coley and N. Pelavas, Gen. Relat. Grav. \textbf{38}, 445 (2006);
V. Pravda, J. Phys. Conf. Ser. \textbf{33}, 463 (2006);
A. Coley, R. Milson, V. Pravda, and A. Pravdova, Class. Quantum Grav. \textbf{21}, L35 (2004);
A. Coley, Class. Quantum Grav. \textbf{25}, 033001 (2008);
V. Pravda, A. Pravdova, and M. Ortaggio, Class. Quantum Grav. \textbf{24}, 4407 (2007).

\bibitem{d5KKD}
K.S. Soh and P.Y. Pac, Phys. Rev. D \textbf{35}, 544 (1987);
A. Macias and H. Dehnen, Class. Quantum Grav. \textbf{8}, 203 (1991).

\bibitem{WuNP}
S.Q. Wu, unpublished notes, paper in preparation.

\bibitem{PJDS}
P.J. De Smet, Gen. Relat. Grav. \textbf{36}, 1501 (2004);
gr-qc/0306026.

\bibitem{BRs}
R. Emparan and H.S. Reall, Phys. Rev. Lett. \textbf{88}, 101101 (2002);
H. Elvang, R. Emparan, D. Mateos, and H.S. Reall, Phys. Rev. Lett. \textbf{93}, 211302 (2004).

\bibitem{Heun}
K. Heun, Math. Ann. \textbf{33}, 161 (1889);
\textit{Heun's Differential Equations}, edited by A. Ronveaux, (Oxford University Press, 1995).

\bibitem{PF}
R. Penrose, Ann. N.Y. Acad. Sci. \textbf{224}, 125 (1973);
R. Floyd, \textit{The dynamics of Kerr fields}, Ph.D Thesis, London University, 1973;
C.D. Collinson, Int. J. Theor. Phys. \textbf{15}, 311 (1976).

\bibitem{KY}
K. Yano, Ann. Math. \textbf{55}, 328 (1952).

\bibitem{TK}
S. Tachibana, T\^{o}hoku Math. J. \textbf{21}, 56 (1969);
T. Kaschiwada, Nat. Sci. Rep. Ochanomizu Univ. \textbf{19}, 67 (1968);
S. Tachibana and T. Kashiwada, J. Math. Soc. Japan \textbf{21}, 259 (1969).

\bibitem{DR}
W. Dietz and R. R\"{u}diger, Proc. Roy. Soc. Lond. A \textbf{375}, 361 (1981).

\bibitem{PenE}
J.N. Goldberg, Phys. Rev. D \textbf{41}, 410 (1990);
E.N. Glass and J. Kress, J. Math. Phys. \textbf{40}, 309 (1999).

\bibitem{Spcm}
M. Tanimoto, Nucl. Phys. B \textbf{442}, 549 (1995);
R.H. Rietdijk and J.W. van Holten, Nucl. Phys. B \textbf{472}, 427 (1996);
M. Cariglia, Class. Quantum Grav. \textbf{21}, 1051 (2004).

\bibitem{PDirac}
R. G\"{u}ven, Phys. Rev. D \textbf{16}, 1706 (1977);
M. Martellini and A. Treves, Phys. Rev. D \textbf{15}, 3060 (1977);
S.M. Wagh and N. Dadhich, Phys. Rev. D \textbf{32}, 1863 (1985);
B.R. Iyer and A. Kumar, Phys. Rev. D \textbf{18}, 4799 (1978);
Pramana, \textbf{11}, 171 (1978);
\textbf{12}, 103 (1979);
D.A. Leahy and W.G. Unruh, Phys. Rev. D \textbf{19}, 3509 (1979);
B. Punsly, Phys. Rev. D \textbf{34}, 1680 (1986);
C.L. Pekeris, Phys. Rev. A \textbf{35}, 14 (1987);
C.L. Pekeris and K. Frankowski, Phys. Rev. A \textbf{39}, 518 (1989);
B.H.J. McKellar, G.J. Stephenson, and M.J. Thomson, J. Phys. A: Math. Gen. \textbf{26}, 3649 (1993);
M. Ahmed and A.K. Mondal, Phys. Lett. A \textbf{184}, 37 (1993);
M. Ahmed, Phys. Lett. B \textbf{258}, 318 (1991);
F. Belgiorno and M. Martellini, Phys. Lett. B \textbf{453}, 17 (1999);
Y.J. Wang and Z.M. Tang, Chin. Phys. \textbf{10}, 475 (2001);
Astrophys. Space Sci. \textbf{281}, 689 (2002);
S.K. Chakrabarti and B. Mukhopadhyay, Mon. Not. R. Astron. Soc. \textbf{317}, 979 (2000);
D. Batic, J. Math. Phys. \textbf{48}, 022502 (2007).

\bibitem{SWH}
S.W. Hawking, Nature (London) {\bf 248}, 30 (1974);
Commun. Math. Phys. {\bf 43}, 199 (1975); \textbf{46}, 206(E) (1976);
H. Nomura, S. Yoshida, M. Tanabe, and K. Maeda, Prog. Theor. Phys. \textbf{114}, 707 (2005).

\bibitem{DNP}
D.N. Page, Phys. Rev. D \textbf{13}, 198 (1976);
\textbf{14}, 3260 (1976);
E. Jung and D.K. Park, Nucl. Phys. B \textbf{731}, 171 (2005).

\bibitem{InSta}
V. Cardoso, \'{O}.J.C. Dias, J.P.S. Lemos, and S. Yoshida, Phys. Rev. D \textbf{70}, 044039 (2004);
\textbf{70}, 049903(E) (2004);
V. Cardoso and \'{O}.J.C. Dias, Phys. Rev. D \textbf{70}, 084011 (2004);
E. Berti, K.D. Kokkotas, and E. Papantonopoulos, Phys. Rev. D \textbf{68}, 064020 (2003);
K. Murata and J. Soda, arXiv:0803.1371 [hep-th].

\bibitem{AFA}
A.N. Aliev and V.P. Frolov, Phys. Rev. D \textbf{69}, 084022 (2004);
A.N. Aliev, Phys. Rev. D \textbf{75}, 084041 (2007);
Class. and Quant. Grav. \textbf{24}, 4669 (2007).

\bibitem{SY}
M. Sakaguchi and Y. Yasui, Int. J. Mod. Phys. \textbf{A 21}, 2331 (2006).

\end{thebibliography}
\end{document}